\documentclass[12pt]{article}
\usepackage{epsfig}
\usepackage{epsfig}
\usepackage{amsmath,graphicx,latexsym}
\def\bbbc{{\mathchoice {\setbox0=\hbox{$\displaystyle\rm C$}\hbox{\hbox
 to0pt{\kern0.4\wd0\vrule height0.9\ht0\hss}\box0}}
 {\setbox0=\hbox{$\textstyle\rm C$}\hbox{\hbox
 to0pt{\kern0.4\wd0\vrule height0.9\ht0\hss}\box0}}
 {\setbox0=\hbox{$\scriptstyle\rm C$}\hbox{\hbox
 to0pt{\kern0.4\wd0\vrule height0.9\ht0\hss}\box0}}
 {\setbox0=\hbox{$\scriptscriptstyle\rm C$}\hbox{\hbox
 to0pt{\kern0.4\wd0\vrule height0.9\ht0\hss}\box0}}}}

\newcommand{\beq}{\begin{equation}}
\newcommand{\eeq}{\end{equation}}
\newcommand{\beqa}{\begin{eqnarray}}
\newcommand{\eeqa}{\end{eqnarray}}
\newcommand{\bagn}{\begin{align}}
\newcommand{\eagn}{\end{align}}
\begin{document}
\setcounter{page}{0}
\thispagestyle{empty}
%
%
%
\markboth{ }{ }
\renewcommand{\baselinestretch}{1.0}\normalsize
\hfill HD-THEP-98-XX \\
\vspace{\baselineskip}
\hfill HD-TVP-98-XX \\
\vspace*{2cm}
\renewcommand{\thefootnote}{\fnsymbol{footnote}}
\begin{center}
{\LARGE\bf Boundary between 
Hadron and Quark/Gluon 
Structure of Nuclei\\ }
\end{center}
\bigskip
\makeatletter \begin{center}
\large H.J.~Pirner and  J. P. Vary  \vspace*{0.3cm} \\
{\it Institut f\"ur Theoretische Physik der Universit\"at Heidelberg, Germany }
{\it Department of Physics and Astronomy, Iowa State University, Ames, IA 50011, USA }

\end{center} \makeatother
{\begin{center} (\today) \end{center}}
\vspace*{2cm}
\begin{abstract}
\noindent

We show that the boundary between quark-dominated and hadron-dominated
regions of nuclear structure may be blurred by 
multi-nucleon quark clusters arising from color percolation.
%
Recent experiments supporting partial percolation in cold nuclei and full percolation
in hot/dense nuclear matter include:
deep inelastic lepton-nucleus scattering, relativistic heavy-ion collisions and 
the binding energy in $^5 He_{\Lambda}$.
%
\end{abstract}

\newpage
\renewcommand{\baselinestretch}{1.1}\normalsize
\renewcommand{\thefigure}{\arabic{figure}}
\renewcommand{\thefootnote}{\arabic{footnote}}
\setcounter{footnote}{0}

\section{Introduction} 
\setcounter{section}{0} 
Heavy ion experiments and lattice-gauge simulations of 
Quantum Chromodynamics (QCD) intensively investigate 
the quark-gluon phase transition relevant to the early universe.  
Another challenge is to understand, 
under varying external conditions,
the roles of fundamental quarks and gluons in hadronic and nuclear matter.  Where are they
essential to replace low-energy descriptions of nuclei based on hadrons
(neutrons, protons and mesons)?
From lattice QCD simulations, we understand quantitatively the masses of the hadrons.  
Similarly, we know that quark and gluon color charges make the strong coupling
constant grow at distances larger than the hadron size and generate their own 
confinement within color-neutral hadrons.  However, a theoretical 
understanding of longer range QCD dynamics, beyond the size of a hadron, 
is still a challenge for lattice simulations. 
We present evidence for an intermediate-range phenomena, 
the formation of multi-quark clusters in cold and hot equilibrated baryonic matter 
that blurs the boundary between the hadron and quark-gluon phases.

We adopt classical percolation theory
\cite { Baym, Satz:1998kg,Pirner:1980eu, Sato:1986vs}
to bridge the gap between quark and hadron physics. In our hybrid
approach we take nuclear matter to be composed of color neutral
clusters, which consist at zero temperature of 3, 6, 9, 12, ... quarks and at
finite temperature we include quark-antiquark ($q \bar q$) states for
the mesons. These larger units have small probabilities in cold nuclei, but are
conducting color, like subdomains of ionized plasma in a electrically
neutral gas. We assume there is a common size parameter for the lowest
mass nucleon and pion which determines the overlap of hadronic states
at finite density and consequently the probabilities to form these
multi-hadron clusters.  We expect the presently neglected gluon
content of these systems to become important at very short distances.

\begin{figure}[h!]
\setlength{\unitlength}{1.cm}
  \begin{center}
\epsfig{file=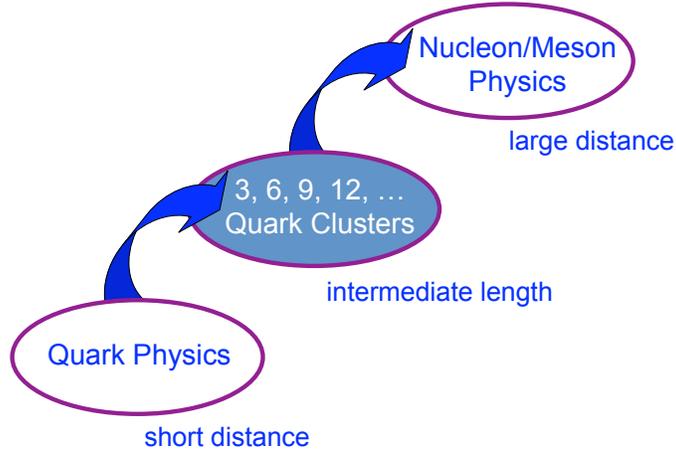, width=10cm} 
  \end{center}
  \caption{(Color online) The separation of quark physics, 
the physics of color neutral multi-quark clusters  and  
and nucleon/meson physics.}
  \label{fig:Trad}
\end{figure}

In fig.1 we summarize the percolation approach to nuclear
physics where, between the  well-established hierarchy of
quark/gluon  physics and nucleon/meson physics, we insert a new layer consisting of
color neutral multi-quark clusters. 
Percolation theory allows us to connect these regions and 
appears sufficient to unify the microscopic and
macroscopic physics with an accuracy of $\approx 10$\%.  Solutions for 
cluster probabilities in nuclei have previously been reported 
\cite{Sato:1986vs,Guttner:1985hc}.  
Current experiments probe nuclei with high resolution and/or high
excitation energies where this quark-clustering layer becomes manifest
as we now discuss.  Future electron-nucleus (e-A) experiments at JLAB with 12 GeV could
enhance the resolution by an order of magnitude and reveal precise
cluster details.

\section {Deep Inelastic Scattering on Nuclei with $x>1$}

For deep inelastic lepton-nucleus scattering, in Born approximation, the Bjorken-variable 
$x=\frac{Q^2}{2 m_N\nu}$ 
($Q$ is the magnitude of the 4-momentum transfer to the target, $\nu$ is the laboratory
energy loss of the lepton, and $m_N$ is the nucleon mass)
measures the fraction of the struck quark's momentum relative to the nucleon momentum. 
Therefore, a constituent of an isolated nucleon carries $x \leq 1$.  
Quark clustering in nuclei leads to a simple prediction for the region $x>1$ where cooperative
effects  are required to provide scattering support.

Following our previous work \cite{Pirner:1980eu}, we assume that the nucleus consists of 3, 6, 9, 12, ...
quark clusters with definite internal structure (i.e.  defined
structure functions) and cluster probabilities are governed by intermediate range correlations 
evaluated with realistic correlated wavefunctions for $A\leq4$ nuclei, then scaled according to average density for heavier nuclei \cite{Sato:1986vs}. 
We omit effects of additional strong density fluctuations due to, for example, alpha-cluster formation in $A > 4$ nuclei.
Two nucleons form a 6-quark cluster when they are separated by less than a critical distance $d_c = 2R_c = 1 fm$, where $R_c$ is a nucleon's critical "color percolation" radius.
A third nucleon with separation less than $d_c$ from the previous two then joins to make a 9-quark 
cluster, etc.  The probability $p_i(A)$ that a quark selected at random in nucleus A originates from an i-quark cluster (for $i=6, 9, 12 . . .$) decreases rapidly with increasing quark number. 
It follows that the e-A structure functions 
are determined by the 6-quark cluster for $1<x<2$ 
and by the 9-quark cluster for $2<x<3$.   
Independent of the cluster structure functions, per nucleon ratios of
deep inelastic cross sections in two nuclei A and B will be
constant in the respective x- intervals and reflect only the
ratio of cluster probabilities of these nuclei, i.e.
for $1<x<2$
\beqa
\frac{(1/A) d \sigma / dx dQ^2 (e A)}{(1/B) d \sigma/ dx dQ^2 (e B)}= \frac {p_6(A)}{p_6(B)}~, 
\eeqa
and for $2<x<3$
\beqa
\frac{(1/A) d \sigma/ dx dQ^2 (e A)}{(1/B) d \sigma/ dx dQ^2 (e B)}= \frac{p_9(A)}{p_9(B)}~,
\eeqa
and so forth for larger clusters.
We have shown how this model successfully describes the inclusive inelastic cross sections on deuterium in the region $1<x<2$ \cite{Yen} when the 6-quark cluster component is included.

\begin{figure}[h!]
\setlength{\unitlength}{1.cm}
  \begin{center}
\epsfig{file=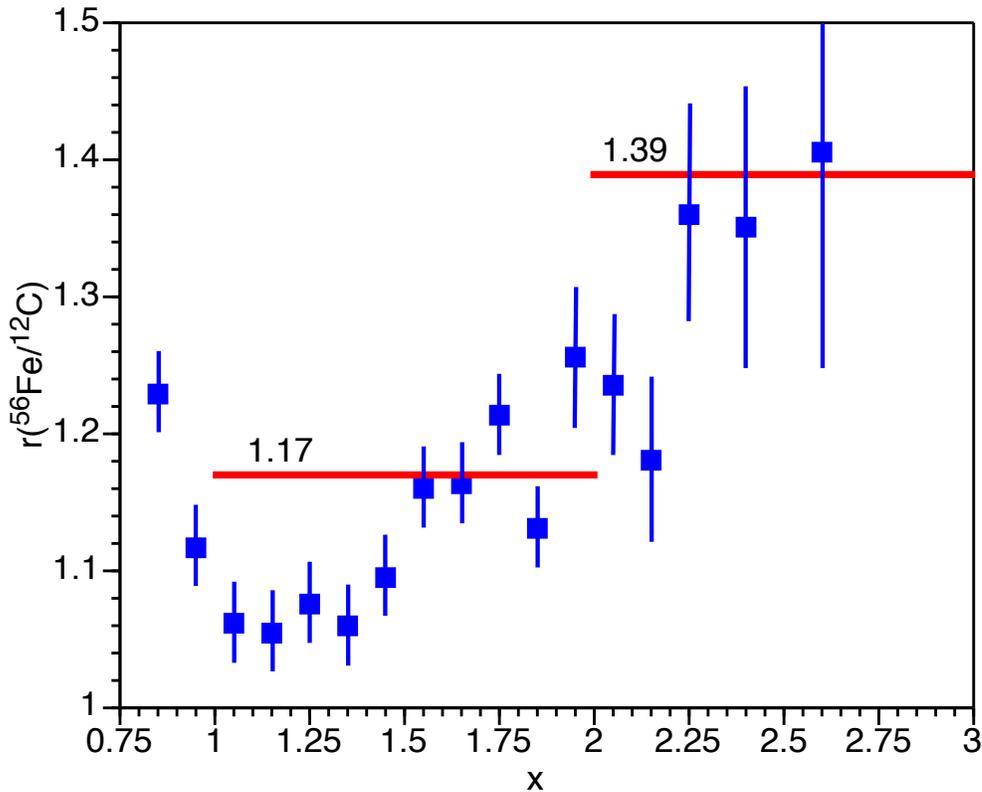, width=15cm}
  \end{center}
  \caption{(Color online) JLAB data for the ratio of per nucleon responses of
  $^{56}$Fe to $^{12}$C over a range of Bjorken $x_B=x$ that exceeds
  unity \cite{Egiyan2006}, the limit for data on an isolated nucleon.  
  We note there is a data point, 
  $r=2.2$ at $x=2.8$, which is far off the vertical scale.
  Multi-quark clusters support a nuclear response above $x=1$.
  The horizontal lines
  show the theoretical predictions (1.17, 1.39) for regions
  dominated by (6,9)-quark clusters respectively\cite{Sato:1986vs}.
  }
  \label{fig:Fe_to_C}
\end{figure}

The intense JLAB electron beam has allowed these  
suppressed regions of phase space to be measured \cite{Egiyan2006}.
In Fig. \ref{fig:Fe_to_C}  we show the measured ratio (discrete points) \cite{Egiyan2006} of per nucleon responses of 
$^{56}$Fe to $^{12}$C 
together with our predictions (horizontal bars)\cite{Sato:1986vs} in the 6-quark cluster region 
$1 <  x <  2$ and the 9-quark cluster region  $2 <  x <  3$. 
We estimate the theoretical ratios have $10 \%$ uncertainty.
In the former kinematic region the experimental ratio for $^{56}$Fe to $^{12}C$ is
$1.17 \pm 0.04 \pm 0.11$\cite{Egiyan2006} in good agreement with our theoretical prediction of
$1.17 \pm 0.12$. 
The experimental ratio above $x=2$ is
quoted\cite{Egiyan2006} as 1.44 with about 33\% uncertainty which agrees
with our theoretical ratio of $1.39 \pm 0.14$.

Although we derive encouragement from the agreement (within large 
uncertainties) for the ratio in the 9-quark cluster region we believe this
agreement may be accidental for two reasons.
First, the large $x$ values in these experiments come with  
rather small energy transfer $\nu$ in the rest system of the nucleus. If one
translates these $\nu$ values into a time resolution then one finds
that this "snapshot" has a rather long exposure time,
namely about $\Delta t= 1/0.3 GeV=0.7 fm/c$. Therefore, the 
short-lived fluctuation of a 9-quark cluster may be represented only
partially in the $2 < x < 3$ cross section. Future e-A
experiments  with new or upgraded facilities
can elucidate the dependence on exposure times. 
They would also allow higher momentum transfer $Q$, which  suppresses the
coherent effects of quasi-elastic quark-cluster knockout \cite{Pirner:1980eu} at
$x=1, 2,$ etc., which may be visible in Fig. \ref{fig:Fe_to_C}.
Second, the calculations\cite{Sato:1986vs} were performed without 
three-nucleon (NNN) interactions
since they were unavailable at that time. NNN correlations 
should reflect the repulsive NNN
force and suppress 9-quark probabilities overall. Due to their intermediate range, we expect
the NNN force will also lead to non-trivial A-dependent effects. These complications
motivate significant theoretical effort to modernize the quark cluster 
probability calculations.

An alternative interpretation\cite{Strikman} of these deep inelastic scattering 
data employs short-range two-nucleon (NN) correlations
to generate very high ($\approx 1 GeV/c$) nucleon relative momenta 
so that  the struck quark may obtain a large momentum and 
exceed the kinematical boundary $x = 1$. 
The high momenta are obtained by strongly repulsive short
range NN-potentials acting at separation distances of $r<0.7-0.8 fm$. 
In so far as this nucleon-based approach generates similar kinematics, 
it may appear to represent a non-relativistic picture dual to our quark-cluster approach.
However, color-conducting multi-quark clusters offer both a relativistic approach and
additional experimental consequences.  For example, virtual photons at high resolution
can peer into the quark cluster zone of color conductivity
and measure details of the quark cluster structure functions.  We may discover, for example, 
that percolating quarks carry less of the total momentum sum rule than
isolated nucleons even though $x > 1$ regions are populated.  This seems plausible on
the basis of elementary physics since the color confining region is enlarged in the 6-quark
cluster \cite{Nachtmann:1983py}. 


A recent experiment probing the $x  < 1$ region suggests
that, while $^3$He, $^4$He and $^{12}$C responses follow  a density scaling trend, 
the $^9$Be response does not\cite{Seely}.
In addition, a detailed study of hadron-nucleus data at $x>1$ shows significant deviations
from simple density scaling for $^6$Li and $^7$Li nuclei\cite{Peterson}.
We understand these Li and Be anomalies as motivating the inclusion of alpha-clustering effects which are expected to be especially significant for these nuclei.  For this reason and our neglect of NNN potentials, we prefer to focus on larger systems that we believe are less sensitive to these corrections.

\section{Relativistic Heavy Ion Collisions}

We now show that the quark cluster description is
appropriate at high excitation energies produced in relativistic heavy ion collisions. 
Cluster probabilities are mainly a function of the local hadron density and,
to proceed to finite temperature, we introduce a pion component. 
We assume that the critical separation distance for the pions 
to transition into larger quark clusters is the
same as for nucleon percolation, $d_c=1 fm$, and that 
$d_c$ is temperature independent.
Therefore we adjust the baryon chemical potential 
so that the sum of the baryon and pion densities is 
the same as the density of baryons at $T=0$ to maintain approximately constant 
color percolation probability.  Then

\begin{equation}
\rho_{hadrons}(\mu,T)=\rho_{baryons}(\mu_b,T)+\rho_{pions}(\mu_{\pi},T).
\end{equation}

We set the pion chemical potential $\mu_{\pi}=0$ at $T=0$ to fix 
the reference baryon chemical potential $\mu_b$ at normal nuclear matter
density $0.17$ baryons/$fm^3$ (Fermi momentum $k_f=1.36~fm^{-1}$).
The nucleon dispersion relation is influenced by the nuclear mean field
potential for which we take \cite{Chen}

\begin{equation}
\epsilon(k)=\sqrt{(k^2+(m_N+V_s(k))^2)}+V_0(k)
\eeq
where the scalar and vector self-energies are momentum dependent
\beqa
V_s(k)&=&-0.35 GeV(1-0.05 (k/k_f)^2)\\
V_0(k)&=&+0.29 GeV(1-0.05 (k/k_f)^2)
\eeqa

\begin{figure}[h!]
\setlength{\unitlength}{1.cm}
  \begin{center}
\epsfig{file=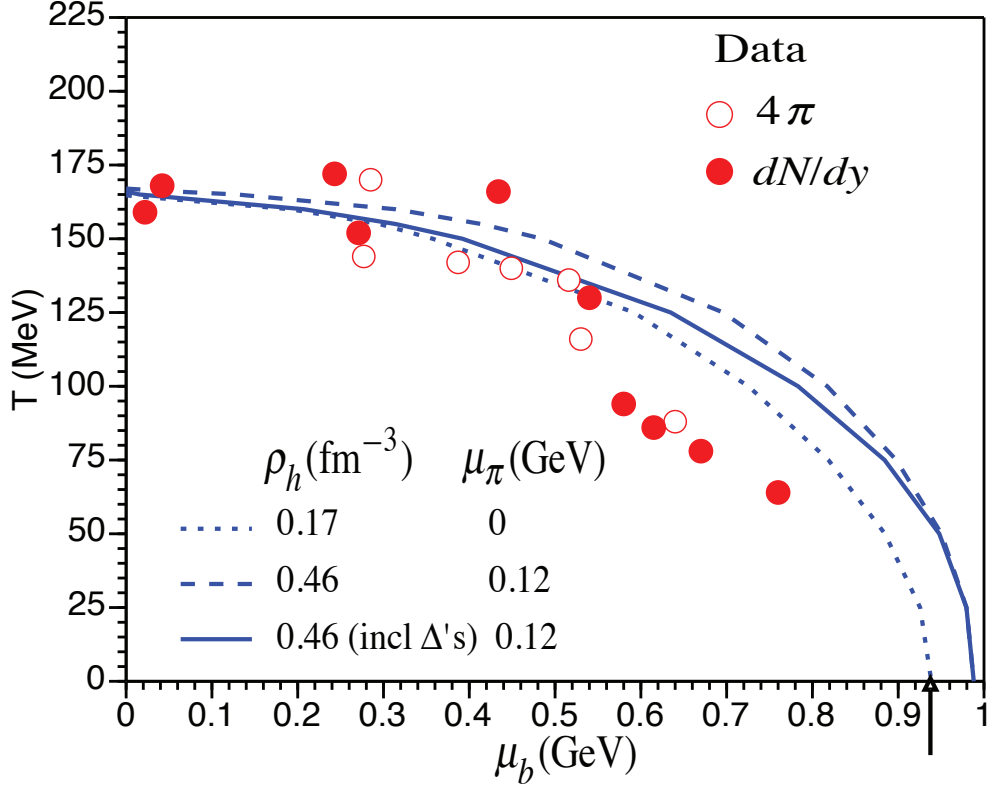, width=15cm} 
  \end{center}
  \caption{(Color online) The boundary between hadron- and quark/gluon-matter
from our quark percolation model for $50\%$ percolation rate (upper curves) and
$22\%$ percolation rate (lower curve) in a temperature T versus baryon chemical potential
$\mu_b$ phase diagram.  The parameters of the three theory curves are
specified in the legend. The arrow indicates $\mu_b$ for nuclear matter
at $T=0$.
The freeze-out data are compiled from different experiments\cite {Andronic:2005yp}.}
  \label{fig:NuclearEOS} 
\end{figure}

The freeze-out data \cite {Andronic:2005yp} extracted from heavy ion
collisions at high temperatures are shown in Fig. 3 
as a function of $T$ and $\mu_b$
along with the theoretical (dotted) curve which corresponds to a mostly
hadronic system of pions and nucleons.
On this curve, each quark is percolating $22\%$ of the time
in $i>3$-quark clusters.
We also plot in Fig. 3 a curve with a higher percolation rate which we call 
"percolating nuclear matter" and find $\mu_b$ 
where more than $50\%$ of all nucleons have transitioned 
to 6, 9, 12, ... -quark clusters at $T=0$.  This $T=0$ state requires 
$\rho_{baryon}=0.46 fm^{-3}$ i.e. 2.7 times nuclear density and therefore
has a larger chemical potential than nuclear matter.
For this case, we also adopted $\mu_{\pi}=120 MeV$, 
used to fit the transverse momentum
dependence of the pion spectra from a kinetically equilibrated pion
distribution\cite {Gerber:1990yb}. 
Taking into account also the $\Delta$- resonance at 1.236 GeV shifts
the higher temperature percolation curve (solid line) towards the curve 
corresponding to nuclear matter density (dotted line). 

The percolation model is a geometrical model and is not sensitive to the
decay properties of resonances, but more to the number of
hadronic states in a certain volume.  
The classical percolation transition is a cross over transition 
in finite systems and the transition itself is not well defined.
Thus quark percolation blurs the boundary between the hadron 
phase and the quark/gluon phase.

\section{ Extended Quark Orbitals in Hypernuclei}

Another avenue to explore quark cluster structures is
to replace one of the nucleons with a strange spectator, a $\Lambda-$hyperon,
producing a hypernucleus.
The successful explanation of the hypernucleus spin-orbit potential 
based on quark dynamics \cite{Pirner:1982kb} motivates 
a closer look at possible quark cluster effects. 
In particular, we focus on a longstanding puzzle \cite
{Hungerford:1984mf} of the sudden decrease of the binding energy in
$^5$He$_{\Lambda}$ compared with the binding energy of a baryon-based
calculation with two-body forces.  The $^4$He nucleus is doubly magic
in the language of the shell model with protons and neutrons but the full 
occupation of the 0s -state by protons and
neutrons does not preclude the addition of a $\Lambda-$hyperon to the
0s -state.  With two-body forces alone one then obtains overbinding of
the A=5 hypernucleus by $1 - 3 MeV$.  One way out is through effective
many-baryon potentials.  Indeed, calculations with purely baryonic
degrees of freedom and a three-body $\Lambda$-NN force
resulting from an explicit $\Sigma$- component in the wave function 
\cite{Nemura:2002fu} seem to show the repulsive
character for the total three-body contributions.  
However, if the three quark substructure of the $\Lambda$
matters, i.e. up(u)+down(d)+strange(s), then for the
12-quark cluster configuration of $^4$He, the 
$\Lambda$'s up and down
quarks cannot occupy the 0s-shell which can
accommodate only 12 u- and d- quarks according to the Pauli principle.
The observation of such extended quark orbitals would be the
best proof for color percolation in nuclei as G. Baym has already
indicated in 1979 \cite {Baym}.

Let us make a simple estimate of this effect.
The mean spacing in the constituent quark shell model
would be larger by a factor three based on the ratio of nucleon to constituent 
quark masses

\beq
\hbar \omega_Q= 3 \hbar \omega_N=75MeV
\eeq

Including a size factor for the smaller 12 quark cluster compared
to the nuclear size could easily boost this mean spacing by
a factor of two.

\begin{figure}[h!]
\setlength{\unitlength}{1.cm}
  \begin{center}
\epsfig{file=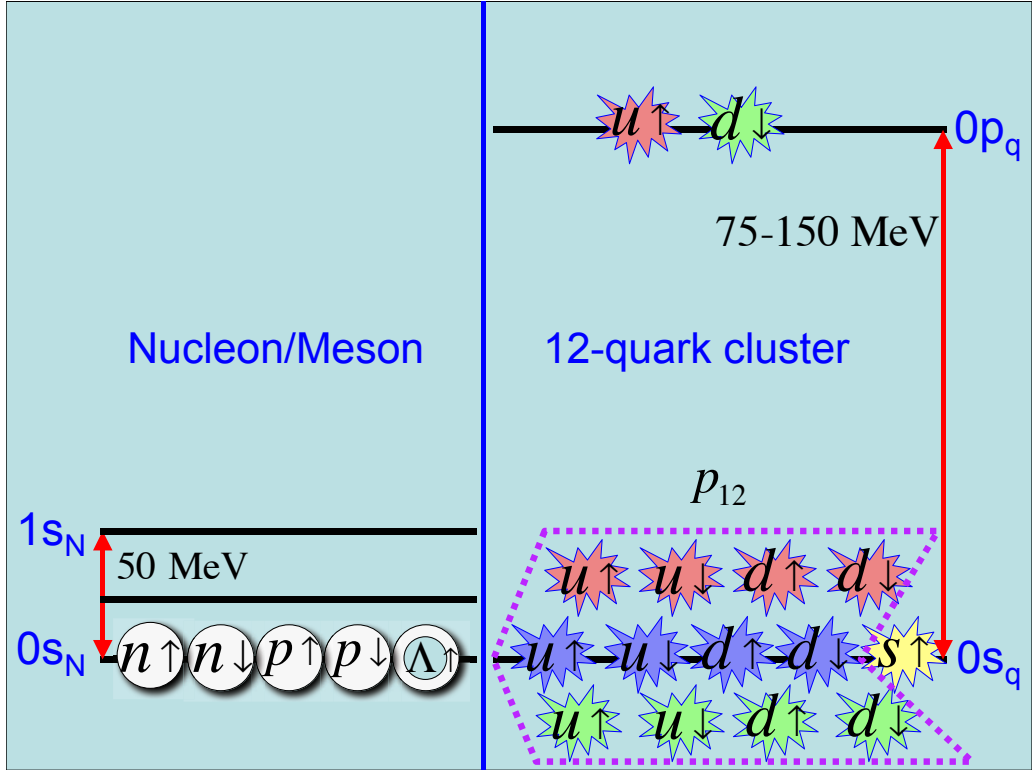, width=15cm} 
  \end{center}
  \caption{(Color online) Sketch of the two possible views of the Lambda attached to the the 
  lowest configuration of two neutrons(n) and two protons(p).  On the left, the Lambda and 
  the nucleons are 
  structureless
  baryons with spin projections
  indicated. On the right, their
  quark substructure is taken into account with spin and color attributes sketched.  
  In the latter case, the 12-quark cluster configuration 
  that appears with probability $p_{12}= 0.007$ in the quark cluster model, forces 
  the up(u) and down(d) quarks of the Lambda to be in the next higher, Pauli-allowed, orbit $0p_q$.
  The Lambda's strange(s) quark may occupy the lowest $0s_q$ level. }
  \label{fig:Lambda-5He} 
\end{figure}

Therefore a 12-quark cluster with two extra u- and d- quarks 
would need an extra excitation energy of approximately
150-300 MeV. Our quark cluster model provides \cite{Sato:1986vs} 
$p_{12}(^4He)=0.007$,
a small probability for such an extended 12-quark cluster in
$^4He$ (Fig. 4). Multiplying this probability with the Pauli-required u- and d-
quark excitations of 150-300 MeV would give a decrease in
$^5He_{\Lambda}$ binding of 
$1-2 MeV$, 
in rough agreement
with experiment. This one case is not sufficient to establish
the existence of extended quark shell model states.  One should look
at heavier hypernuclei where the additional
$\Lambda$ has to split up in a low lying s-quark and an excited
ud-pair.  Future precision experiments could explore the binding
energy systematices of heavier closed shell systems such as:
$^{16}O_{\Lambda},^{17}O_{\Lambda}$ and
$^{16}N_{\Lambda},^{17}N_{\Lambda}$. 

Additional experimental tests of quark clustering can be performed such as
measuring high mass di-lepton pair production from nuclear targets\cite{Hari1986}
(Drell-Yan process).

If one asks whether our understanding of QCD is complete,
the question may be recast to whether additional theoretical or 
experimental work can improve the situation decisively.
Compressed baryonic matter \cite {Compressed} is an area where our theoretical 
and experimental understanding is not as well developed as baryon-free matter.
For nuclei a theory which contains nucleons and multi-quark cluster states is
an efficient way to describe the transition from purely
nucleonic matter to quark matter at high baryon density.
A hybrid model 
with quark clusters would define a major advance
if one would be able to rewrite  the smooth infrared limit 
of lattice QCD in a field theoretic continuum picture. Such a picture should also 
continue the phase diagram into the totally percolated phase where 
perhaps the formation of a $qq$ -dimer condensate \cite{Rezaeian:2006yj} and BCS-pairing 
\cite{Schafer:1999jg} of quarks can be found in  analogy to the physics of cold atoms. 
Quark percolation and these other features of the QCD phase diagram 
would also impact astrophysics.

{\bf Acknowlegments} \\
We thank V. Burkert for providing the JLAB experimental data\cite{Egiyan2006}.  We acknowledge valuable discussions with V. Burkert, W. Brooks, S. Coon, J. Knoll, R.J. Peterson, J. Stachel and B. Povh. This work was supported in part by USDOE grant DE-FG-02-87ER40371 and, in part, by the Alexander von Humboldt Foundation.

\end{document}